\documentclass[aps,floatfix,superscriptaddress,reprint,10pt,pra]{revtex4-1}
\usepackage{bbm}
\usepackage{bm}
\usepackage{amsmath}
\usepackage{amssymb}
\usepackage{multirow}
\usepackage{empheq}
\usepackage{graphicx}
\usepackage{mathrsfs}
\usepackage{amsfonts}
\usepackage{amsthm}
\usepackage{color}
\usepackage{bigints}
\usepackage{txfonts}
\usepackage{hyperref}
\usepackage{comment}
\hypersetup{
     unicode=false,          
     pdftoolbar=true,        
     pdfmenubar=true,        
     pdffitwindow=false,     
     pdfstartview={FitH},    
     pdftitle={My title},    
     pdfauthor={Author},     
     pdfsubject={Subject},   
     pdfcreator={Creator},   
     pdfproducer={Producer}, 
     pdfkeywords={keyword1} {key2} {key3}, 
     pdfnewwindow=true,      
     colorlinks=false,       
     linkcolor=red,          
     citecolor=green,        
     filecolor=magenta,      
     urlcolor=cyan           
}
\makeatletter \tolerance = 10000 \tolerance = 10000
\usepackage{color}

\definecolor{darkblue}{rgb}{0.,0.,0.4}
\definecolor{darkred}{rgb}{0.5,0.,0.}
\definecolor{BlueViolet}{RGB}{138,43,226}
\definecolor{SkyBlue}{RGB}{30,144,255}
\definecolor{DarkGreen}{RGB}{0,100,0}

\usepackage[normalem]{ulem}

\makeatother
\begin{document}
\title{
Helicity controlled spin Hall angle in the 2D Rashba altermagnets
}

\author{Weiwei Chen}

\affiliation{Key Laboratory of Intelligent Manufacturing Quality Big Data Tracing and Analysis of Zhejiang Province, College of Sciences, China Jiliang University, Hangzhou, 310018, China}

\author{Longhai Zeng}

\affiliation{Key Laboratory of Intelligent Manufacturing Quality Big Data Tracing and Analysis of Zhejiang Province, College of Science, China Jiliang University, Hangzhou, 310018, China}

\author{W. Zhu}
\email[Corresponding author: ]{zhuwei@westlake.edu.cn}
\affiliation{ School of Science, Westlake University, Hangzhou 310024, China}

\date{\today}
\begin{abstract}
We investigate the efficiency of charge-to-spin conversion in two-dimensional Rashba altermagnets, a class of materials that merge characteristics of both ferromagnets and antiferromagnets. Utilizing quantum linear response theory, we quantify the longitudinal and spin Hall conductivities in this system and demonstrate that a substantial enhancement of the spin Hall angle is achieved below the band crossing point through the dual effects of relativistic spin-orbit interaction and nonrelativistic altermagnetic exchange interaction. Additionally, we find that skew scattering and topology-related intrinsic mechanisms are almost negligible in this system, which contrasts with conventional ferromagnetic Rashba systems. Our findings not only advance the understanding of spin dynamics in Rashba altermagnets but also pave the way for novel strategies in manipulating charge-to-spin conversion via the sophisticated control of noncollinear in-plane and collinear out-of-plane spin textures.
\end{abstract}
\maketitle

\textit{Introduction.---}
Altermagnets represent an emerging class of magnetic materials exhibiting hybrid properties, bridging the gap between traditional ferromagnets and antiferromagnets \cite{Smejkal2022prx1,Smejkal2022prx2,Smejkal2022nrm,Bai2023prl,Krempasky2024nature,Smejkal2020sa,Ouassou2023prl,Osumi2024prb}. This emerging material category has attracted significant attention for its promising applications in spintronics, particularly in conversion between charge and spin \cite{Naka2019nc,Naka2021prb,Ma2021nc,Hernandez2021prl,Li2024arxiv,Bai2022prl}. Notable experimental observations include out-of-plane spin torque induced by in-plane charge current without an external magnetic field \cite{Li2024arxiv}, spin-to-charge conversion stemming from altermagnetic spin splitting effect \cite{Bai2023prl}, and spin-current generation arising from anisotropically spin-split bands \cite{Hernandez2021prl,Bai2022prl}. Exploring how to improve the efficiency of these conversion processes is therefore a highly valuable research endeavor.

Previous research on altermagnets has primarily focused on the anisotropic antiparallel spin states \cite{Karube2022prl,Ma2021nc}; however, the spin-orbit coupling (SOC) effect existing in most altermagnetic materials can substantially alter the spin directions of quasiparticles \cite{Reimers2024nc,Reichlova2024nc,Sasabe2023prl}.
The interplay between nonrelativistic collinear spin splitting, driven by altermagnetic exchange interaction, and relativistic noncollinear spin splitting, induced by spin-orbit coupling, greatly enriches the spin textures in the crystal momentum space \cite{Smejkal2020sa}. The role of spin texture in quantum transport is crucial, as it fundamentally shapes the electronic properties of materials and creates a fertile ground for engineering the spin-dependent scattering processes,  essential for advanced applications in spintronics \cite{Seo2021cp,Hua2024prb,Gorini2008prb,Chen2024prb}. By manipulating these spin textures, it is possible to selectively enhance or suppress specific transport channels, thereby enabling precise control over the spin Hall effect and enhancement of the efficiency of spin current generation \cite{Smejkal2022prx3,Cui2023prb}.

In this work, we focus on the charge and spin transport in the two-dimensional altermagnet with Rashba-type SOC. While the Rashba SOC induces helical in-plane spin textures \cite{Brosco2016prl}, the altermagnetic exchange interaction drives anisotropic out-of-plane spin polarization. Through the calculation of longitudinal and spin Hall conductivity within the framework of quantum linear theory, we find that the ladder-type vertex correction, which corresponds to the side-jump scattering, has opposite effect to the spin Hall angle below and above the Dirac point. 

Our findings reveal that the side-jump mechanism significantly influences the spin Hall angle, depending on the Fermi level relative to the band crossing point, also called Dirac point. When the Fermi level is above the Dirac point, the polarization direction of out-of-plane spin in backward scattering states is parallel to that in forward scattering states; however, these directions are antiparallel when the Fermi level is below the Dirac point. Consequently, suppressing backward scattering by the side-jump mechanism below the Dirac point not only increases the distribution of transport states but also enhances the spin polarization. This dual enhancement effect significantly boosts spin Hall conductivity, leading to a pronounced increase in the spin Hall angle.
Our results not only advance the understanding of spin dynamics in Rashba altermagnets but also suggest practical strategies for improving spintronic device functionalities through tailored spin texture manipulations.

\textit{Model.---}
The low-energy effective Hamiltonian for a two dimenstional altermagnet subject to Rashba SOC is described as follows \cite{Smejkal2022prx2,Smejkal2022nrm,Ouassou2023prl}, 
\begin{equation}\label{eq:Hamiltonian}
	H=tk^2+2t_Jk_xk_y\sigma_z+\lambda(k_x\sigma_y-k_y\sigma_x).
\end{equation}
Here, $\bm{k}=(k_x,k_y)$ represents the planar momentum, and $\sigma_i$ ($i=x,y,z$) are the Pauli matrices. The first term represents conventional kinetic energy hopping, with parameter $t$ serving as the energy unit in subsequent calculations. The second term reflects anisotropic exchange interaction in the altermagnetic state, characterized by the strength parameter $t_J$. The final term describes Rashba SOC, with a strength of $\lambda$. The quasiparticle state reads
\begin{equation}
	|\bm{k}+\rangle=\left(\begin{array}{c}
		\cos\frac{\theta}{2}\\\sin\frac{\theta}{2}e^{i\phi}
	\end{array}\right)
	\ \ \ \ \text{and}\ \ \ \
	|\bm{k}-\rangle=\left(\begin{array}{c}
		\sin\frac{\theta}{2}\\-\cos\frac{\theta}{2}e^{i\phi}
	\end{array}\right),
\end{equation}
with dispersion relation denoted by $E_{\bm{k}\gamma}=tk^2+\gamma d$, where $(d,\theta,\phi)$ are the spherical coorditates of the vector $\bm{d}=(-\lambda k_y,\lambda k_x,2t_Jk_xk_y)$ and the index $\gamma=\pm1$ distinguishes quasiparticle states at two different subbands. 

\begin{figure}
	\centering
	\includegraphics[width=0.95\linewidth]{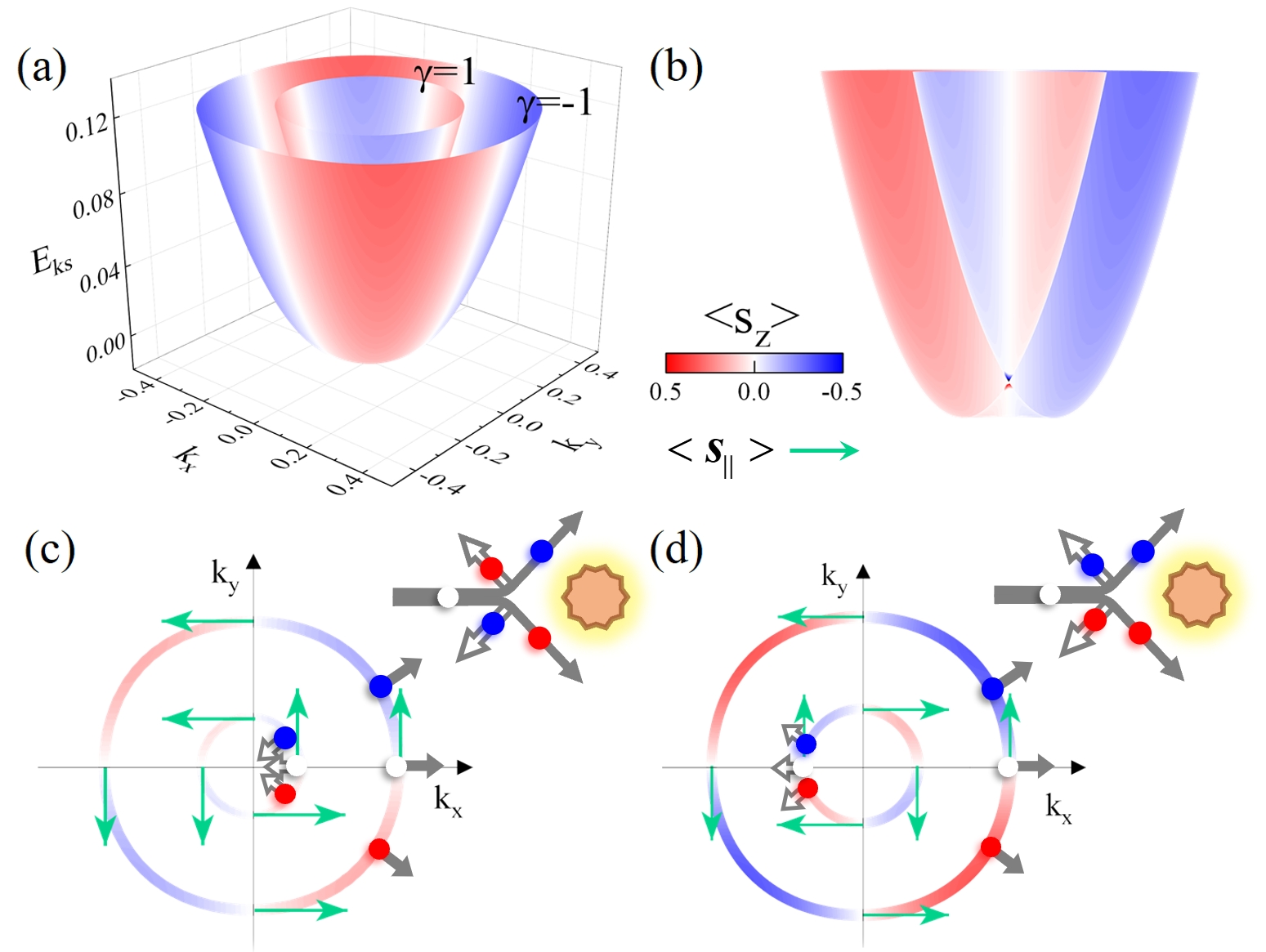}
	\caption{(Color online) (a) Dispersion relation of the Hamiltonian Eq.~(\ref{eq:Hamiltonian}) with parameters: $t_J=0.2$ and $\lambda=0.2$. (b) 2D projection of the dispersion relation at $k_y=0$. The colormap indicates the value of the out-of-plane spin, $\langle s_z\rangle$. (c) and (d) Schematics of the Fermi-surface cut below and above the Dirac point, respectively. Green arrows denote the in-plane spin texture of the Fermi-surface states, $\langle s_{\parallel}\rangle=(\langle s_x\rangle,\langle s_y\rangle)$, which satisfies $\langle s_x\rangle^2+\langle s_y\rangle^2+\langle s_z\rangle^2=\hbar/2$. Gray arrows denote the particle velocity, with solid arrows corresponding to forward scattering states and hollow arrows to backward scattering states. Insets in (e) and (f) depict schematic diagrams of forward and backward scattering processes in two cases. White, red and blue solid circles represent quasiparticle states with $\langle s_z\rangle=0$, $>0$ and $<0$.}
	\label{fig:band}
\end{figure}

Figure~\ref{fig:band} displays the dispersion relations and spin textures of the effective model Eq.~(\ref{eq:Hamiltonian}). The spin texture of each quasiparticle states $|\bm{k}\lambda\rangle$ is quantified by $\langle s_i\rangle=\langle\bm{k}\gamma|\sigma_i|\bm{k}\gamma\rangle\hbar/2$. The two subbands ($\gamma=\pm1$) intersect at $E=0$, defining the Dirac point, around which the Fermi surface splits into two distinct Fermi circles.  In this case, both the in-plane and out-of-plane spin degeneracies are lifted. Specifically, the in-plane spin component $\langle s_{\parallel}\rangle$ exhibits helicity induced by SOC, and the out-of-plane component $\langle s_z\rangle$ shows pronounced anisotropic spin polarization, a manifestation of altermagnetism.

The charge velocities along $x$ and $y$ direction of Hamiltonian Eq.~(\ref{eq:Hamiltonian}) are given by
\begin{equation}\begin{aligned}\label{eq:vertex-charge}
		v_x=&2tk_x+2t_Jk_y\sigma_z+\lambda\sigma_y,\\
		v_y=&2tk_y+2t_Jk_x\sigma_z-\lambda\sigma_x.
\end{aligned}\end{equation}
The corresponding spin velocities, which refer to the effective velocity at which the spin state contributes to the transport properties of a material, are defined as
\begin{equation}\begin{aligned}\label{eq:vertex-spin}
		v_{s,x}=&\frac{1}{2}\{v_x,\sigma_z\}=2tk_x\sigma_z+2t_Jk_y\sigma_0,\\
		v_{s,y}=&\frac{1}{2}\{v_y,\sigma_z\}=2tk_y\sigma_z+2t_Jk_x\sigma_0.
\end{aligned}\end{equation}

Disorder potential is modeled by randomly located $\delta$-function scatterers, $V(\bm{r})=\sum_iV_i\delta(\bm{r}-\bm{R}_i)$, where the average scattering strengths are characterized by $\langle V_i\rangle=0$, $\langle V_i^2\rangle=V_0^2$, and $\langle V_i^3\rangle=V_1^3$. The second and third moments, $\langle V_i^2\rangle$ and $\langle V_i^3\rangle$, are crucial for quantifying the contributions of side-jump and skew scattering mechanisms \cite{Sinova2015rmp}.

\textit{Methods for Lingitudinal and Spin Hall Conductivities.---}
Within the framework of quantum linear response theory, the longitudinal conductivity $\sigma^l_{xx}$ and spin Hall conductivity $\sigma^s_{yx}$ of electrons in solid can be determined using the Kubo formalism,
\begin{equation}
	\sigma^{l/s}_{ij}={\rm Re}(\sigma_{ij}^{l/s,Ia}+\sigma_{ij}^{l/s,Ib}+\sigma_{ij}^{l/s,II})
\end{equation}
with
\begin{equation}\begin{aligned}\label{eq:Kubo-initial}
	\sigma_{ij}^{l/s,Ia}=&\frac{e^2}{h}\int\frac{d^2\bm{k}}{(2\pi)^2}{\rm Tr}[v_iG^R_{\bm{k}}(E_F)v_jG^A_{\bm{k}}(E_F)]_c\\
	\sigma_{ij}^{l/s,Ib}=&-\frac{e^2}{h}\int\frac{d^2\bm{k}}{(2\pi)^2}{\rm Tr}[v_iG^R_{\bm{k}}(E_F)v_jG^R_{\bm{k}}(E_F)]_c\\
	\sigma_{ij}^{l/s,II}=&\frac{e^2}{h}\int dE f(E)\int\frac{d^2\bm{k}}{(2\pi)^2}{\rm Tr}[v_iG^R_{\bm{k}}(E)v_j\frac{dG^R_{\bm{k}}(E)}{dE}\\
	&-v_i\frac{dG^R_{\bm{k}}(E)}{dE}v_jG_{\bm{k}}^R(E)]_c\\
\end{aligned}\end{equation}
Here, $f(E)=\frac{1}{\exp[(E-E_F)/k_BT]+1}$ is the Fermi-Dirac distribution function, ${\rm Tr}$ means the trace in the spin subspace, and the subscript $c$ indicates a disorder configuration average. $G^{R/A}_{\bm{k}}(E)=1/(E-H-\Sigma^{R/A})$ denote the retarded ($R$) and advanced ($A$) Green's functions of the disordered system with self-energy function $\Sigma^{R/A}$. For the longitudinal conductivity ($\sigma^l_{xx}$), the vertices in the Kubo formula are set as $v_i=v_j=v_x$, while for the spin Hall conductivity ($\sigma^s_{yx}$), the vertices are set as $v_i=v_{s,y}$ and $v_j=v_x$. Clearly, $\sigma_{xx}^{l,II}$, arising from the Fermi sea states, does not contribute to the longitudinal conductivity. 

The Green's functions in Eq.~(\ref{eq:Kubo-initial}) can be expressed in the quasiparticle-state basis as $G^{R/A}_{\bm{k}}(E)={\rm diag}[g^{R/A}_{+}(E),g^{R/A}_{-}(E)]$, where $g^{R/A}_{\pm}(E)$ is the retarded/advanced Green's function for quasiparticle state with subband index $\gamma=\pm$. Given the $\delta$-function disorder, the self-energy functions, calculated using the self-consistent Born's approximation (SCBA), are independent of both momentum and the subband index $\gamma$. The real part of self-energy is neglected, as it can be absorbed into the shifted energy levels. Subsequently, the quasiparticle-state Green's function simplifies to $g^{R/A}_{\gamma}(E)=1/(E-E_{\bm{k}\gamma}\pm i/2\tau)$, where $\tau=-\hbar/2{\rm Im}\Sigma^R$ denotes the quasiparticle relaxation time. 

\begin{figure}
	\flushleft
	\includegraphics[width=0.95\linewidth]{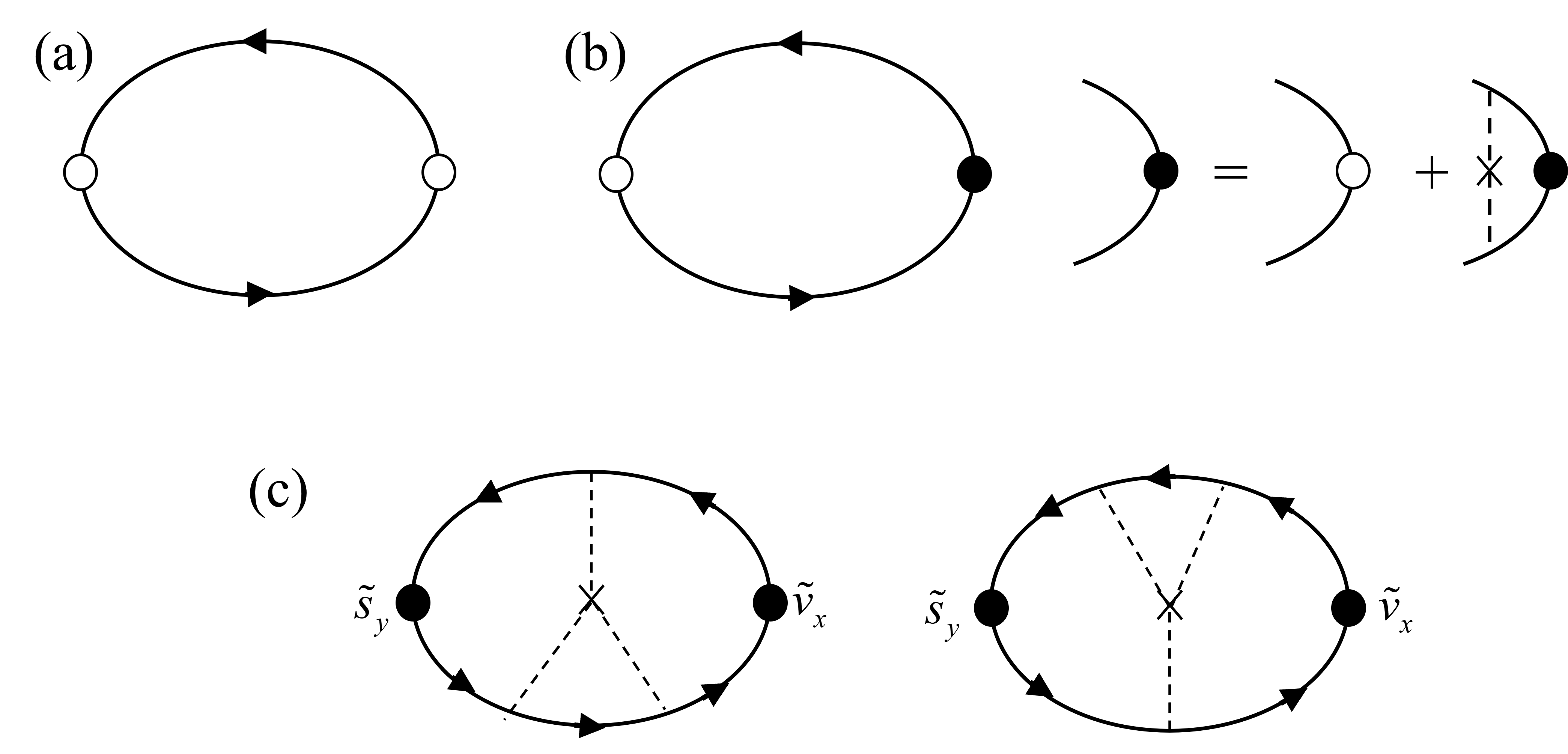}
	\caption{Feynman diagrams illustrating mechanisms contributing to longitudinal and spin Hall conductivity: (a) A bubble diagram representing the intrinsic mechanism from Fermi states. (b) A ladder-type vertex correction, indicative of the side-jump scattering mechanism. (c) A third-order single impurity vertex with current vertices modified by ladder-type vertex correction, depicting skew scattering. \cite{Tse2006prb,Nunner2008prl} }
	\label{fig:feynman-diagram}
\end{figure}

Three primary microscopic mechanisms---intrinsic, side-jump scattering, and skew scattering---dominate the evaluation of the spin Hall effect \cite{Sinova2015rmp,Tse2006prb}. The corresponding Feynman diagrams, derived from the expansion of the Kubo formalism, are illustrated in Fig.~\ref{fig:feynman-diagram} \cite{Tse2006prb,Borunda2007prl,Nunner2007prb,Nunner2008prl,Kovalev2009prb,Yang2011prb}.

\begin{figure}
	\flushleft
	\includegraphics[width=0.95\linewidth]{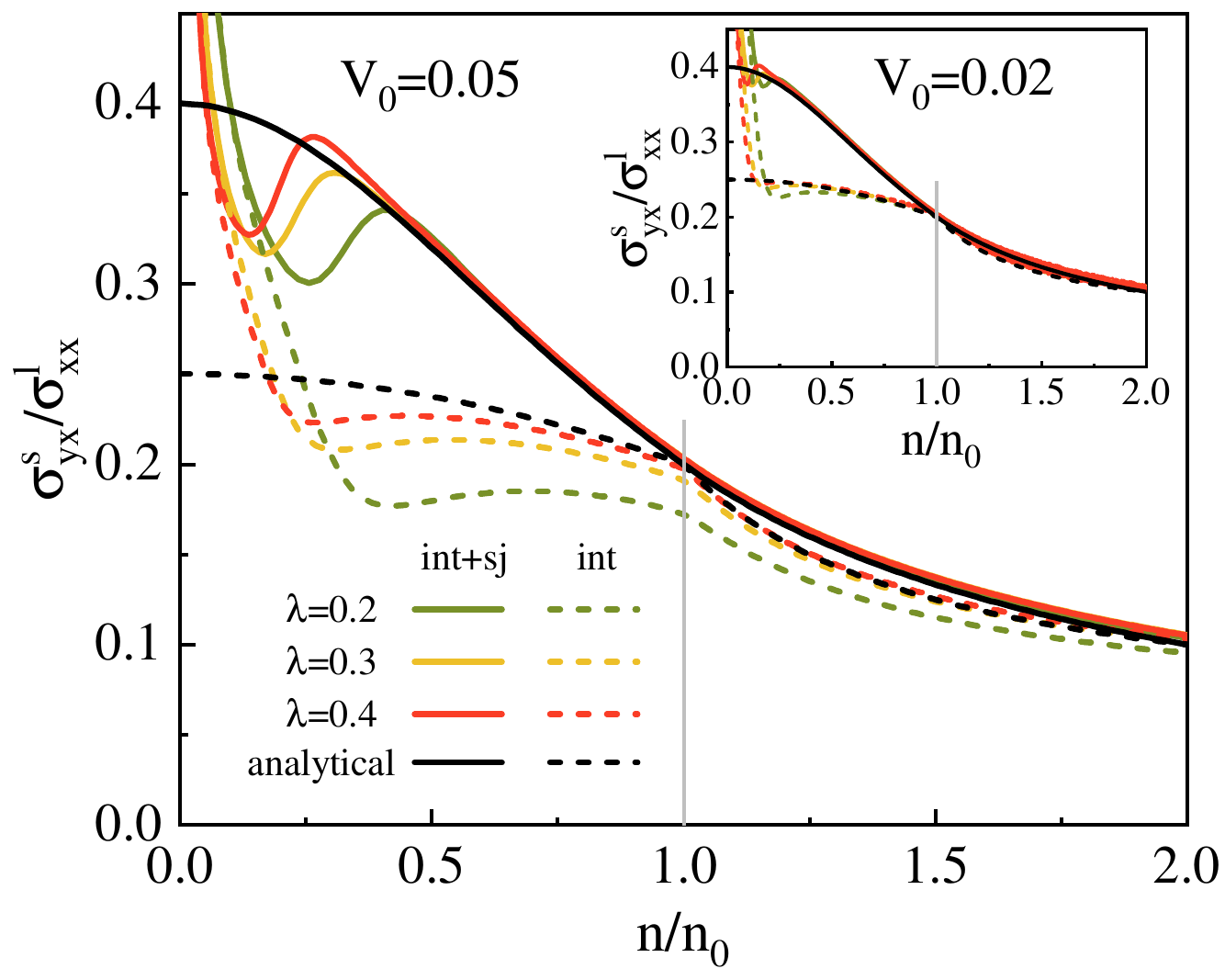}
	\caption{(Color online) Numerical and analytical results of spin Hall angle of the 2D Rashba altermagnet around Dirac point with different spin-orbit coupling strengths $\lambda$. Dashed lines indicate contributions solely from the intrinsic mechanism, while solid lines incorporate both the intrinsic and side jump mechanisms. Analytical solutions, depicted by black dashed and solid lines, are obtained from Eqs.~(\ref{eq:analytic-sxx}-\ref{eq:analytic-syx}) and (\ref{eq:analytic-sxx-sj}-\ref{eq:analytic-syx-sj}). Other parameters: $t_J = 0.2$ and $V_0 = 0.05$ (main panel), $0.02$ (inset). The charge density ratio corresponding to the Dirac point ($n/n_0=1$) is indicated by a gray line.}
	\label{fig:conductivity}
\end{figure}

\textit{Intrinsic mechanism.---} The intrinsic mechanism contributing to the longitudinal conductivity ($\sigma^{l,\text{int}}_{xx}$) and the spin Hall conductivity ($\sigma^{s,\text{int}}_{yx}$) from Fermi states are represented by the bubble Feynman diagram, as shown in Fig.~\ref{fig:feynman-diagram}(a). For $\sigma^{s,\text{int}}_{yx}$, the intrinsic contribution from the Fermi sea states, i.e. $\sigma_{yx}^{s,II}$, should also be included. Numerical results for the spin Hall angle contributed by the intrinsic mechanism, denoted as $\sigma^{s,\text{int}}_{yx}/\sigma^{l,\text{int}}_{xx}$, are presented in Fig.~{\ref{fig:conductivity}} as a function of the charge density ratio $n/n_0$, where $n$ and $n_0$ are the charge densities at the Fermi level and Dirac point, respectively.

Focusing on the region around the Dirac point ($k\ll\frac{\lambda}{t_J}$) and under conditions of weak disorder ($1/\tau\to0$) condition, the Kubo formula can be further simplified. Meanwhile, the $RR$ terms ($g^{R}g^{R}$), interband coherences ($g_{+}g_{-}$, $g_{-}g_{+}$) and the Fermi-sea contribution ($\sigma^{s,II}_{yx}$) are neglected. Thus, $\sigma^{l,\text{int}}_{xx}$ and $\sigma^{s,\text{int}}_{yx}$ can be approximately captured by \begin{equation}\begin{aligned}\label{eq:analytic-intrinsic}
		\sigma^{l,\text{int}}_{xx}\approx&\frac{\pi e^2\tau}{h}\sum_{\gamma=\pm1}\int\frac{d^2\bm{k}}{(2\pi)^2}(v_x^{\gamma\gamma})^2\delta(E-E_{\bm{k}\gamma})\\
		\sigma^{s,\text{int}}_{yx}\approx&\frac{\pi e^2\tau}{h}\sum_{\gamma=\pm1}\int\frac{d^2\bm{k}}{(2\pi)^2}v_{s,y}^{\gamma\gamma}v_x^{\gamma\gamma}\delta(E-E_{\bm{k}\gamma})
\end{aligned}\end{equation}
where
\begin{equation}\begin{aligned}\label{eq:analytic-vx-sy}
		v_x^{\gamma\gamma}=&\langle\bm{k}\gamma|v_x|\bm{k}\gamma\rangle\approx2tk_x+\gamma\frac{\lambda k_x}{k}\\
		v_{s,y}^{\gamma\gamma}=&\langle\bm{k}\gamma|v_{s,y}|\bm{k}\gamma\rangle\approx2t_Jk_x+\gamma\frac{4tt_Jk_y^2k_x}{\lambda k}
\end{aligned}\end{equation} 
are the diagonal components of charge and spin velocities in Eqs.~(\ref{eq:vertex-charge}) and (\ref{eq:vertex-spin}). Plugging Eq.~(\ref{eq:analytic-vx-sy} into Eq.~(\ref{eq:analytic-intrinsic}), we get
\begin{equation}\label{eq:analytic-sxx}
	\sigma^{l,\text{int}}_{xx}
	\approx\frac{e^2}{h}\frac{\lambda^2}{n_iV_0^2}\left\{\begin{array}{ll}
		(\frac{n}{n_0})^2;&n<n_0\\
		2(\frac{n}{n_0})-1;&n\ge n_0
	\end{array}\right.
\end{equation}
and
\begin{equation}\label{eq:analytic-syx}
	\sigma^{s,\text{int}}_{yx}
	\approx\frac{e^2}{h}\frac{\lambda^2}{n_iV_0^2}\frac{t_J}{t}\left\{
	\begin{array}{ll}
		\frac{5}{4}(\frac{n}{n_0})^2-\frac{1}{4}(\frac{n}{n_0})^4;&n<n_0\\
		\frac{1}{2}(\frac{n}{n_0})+\frac{1}{2};&n\ge n_0
	\end{array}\right.
\end{equation}
Here, the relaxation time $\tau$ is approximately obtained from the first-order Born's approximation,
$
	\tau=\frac{2t}{V_0^2}\left\{\begin{array}{ll}
		n/n_0;&n<n_0\\
		1;&n\ge n_0
	\end{array}\right.,
$
The charge density $n$ is obtained in the pure system,
$
	n=\frac{\lambda^2}{2\pi t^2}\left\{\begin{array}{ll}
		\sqrt{1+4tE_F/\lambda^2};&E_F<0\\
		1+2tE_F/\lambda^2;&E_F\ge0
\end{array}\right..
$
It is noteworthy that in the absence of SOC ($\lambda=0$), the Dirac point corresponds to the band edge, as shown in Fig.~\ref{fig:band}, indicating that $n_0=0$. Thus, the small momentum limit approximation $k\ll\frac{\lambda}{t_J}$ is inapplicable. In this case, the approximate analytical solutions for $\sigma^{l}_{xx}$ and $\sigma^{s}_{yx}$ are given by
\begin{equation}\label{eq:sigma-lambda0}
	\sigma^{l}_{xx}(\lambda=0)=t\frac{4\pi ne^2\tau}{h},
	\ \ \text{and}\ \ 
	\sigma^{s}_{yx}(\lambda=0)=t_J\frac{4\pi ne^2\tau}{h}
\end{equation}
which differ from the results in Eqs.~(\ref{eq:analytic-sxx}) and (\ref{eq:analytic-syx}) with $\lambda=0$, resulting in a spin Hall angle of $\sigma^{l}_{xx}/\sigma^{s}_{yx}=t_J/t$.


\textit{Side-jump scattering.---}
The side-jump scattering contribution, corresponding to the ladder-type vertex correction in the Feynman diagrams as shown in Fig.~\ref{fig:feynman-diagram}(b), can be obtained by solving the Bethe-Salpeter equation
\begin{equation}\label{eq:Bethe-Salpeper-1}
	\tilde{v}_x^{LM}(\bm{k},E)=v_x(\bm{k})+V_0^2\int\frac{d^2\bm{k}'}{(2\pi)^2}G^{L}_{\bm{k}'}(E)\tilde{v}_x^{LM}(\bm{k}',E)G^{M}_{\bm{k}'}(E)
\end{equation}
where the subscripts $L,\ M=R,\ A$ indicate retarded or advanced quantities. The resolved dressed vertex is 
\begin{equation}\label{eq:vertex-correction-charge}
	\tilde{v}^{LM}_x=2tk_x+2t_Jk_y\sigma_z+\lambda^{LM}_c\sigma_y
	\ \ \text{with} \ \
	\lambda_c^{LM}=\frac{1+q^{LM}_1}{1-q^{LM}_2}\lambda
\end{equation}
where
\begin{equation}\begin{aligned}
		q^{LM}_1=&\frac{V_0^2}{4}\int\frac{d^2\bm{k}}{(2\pi)^2}[\frac{4tk^2_x}{d}(g^L_+g^M_+-g^L_-g^M_-)\\
		&+\frac{8t^2_Jk^2_xk^2_y}{d^2}(g_+^L-g_-^L)(g_+^M-g_-^M)+i\frac{4t_Jk^2_y}{d}(g^L_+g^M_--g^L_-g^M_+)]\\
		q^{LM}_2=&\frac{V_0^2}{4}\int\frac{d^2\bm{k}}{(2\pi)^2}[(g_+^L+g_-^L)(g_+^M+g_-^M)\\
		&-\frac{4t^2_Jk^2_xk^2_y}{d^2}(g_+^L-g_-^L)(g_+^M-g_-^M)]
\end{aligned}\end{equation}
Above, ladder-type correction is applied to the charge vertex, modifying $v_x$ to $\tilde{v}_x$. Equivalently, we can also apply ladder-type correction to the spin vertex, modifying $v_{s,y}$ to $\tilde{v}_{s,y}$. Similarly, using the Bethe-Salpeter equation, we resolve the dressed spin vertex as
\begin{equation}\label{eq:vertex-correction-spin}
	\tilde{v}_{s,y}^{LM}=2t_Jk_x+2tk_y\sigma_z+\lambda_s^{LM}\sigma_y,
	\ \ \text{with} \ \
	\lambda_s^{LM}=\frac{q_{s1}^{LM}}{1-q^{LM}_{s2}}\lambda
\end{equation}
where
\begin{equation}\begin{aligned}
		q^{LM}_{s1}=&\frac{V_0^2}{4}\int\frac{d^2\bm{k}}{(2\pi)^2}[\frac{4t_Jk^2_x}{d}(g^L_+g^M_+-g^L_-g^M_-)\\
		&+\frac{8tt_Jk^2_xk^2_y}{d^2}(g_+^L-g_-^L)(g_+^M-g_-^M)+i\frac{4tk^2_y}{d}(g^L_+g^M_--g^L_-g^M_+)]\\
		q^{LM}_{s2}=&\frac{V_0^2}{4}\int\frac{d^2\bm{k}}{(2\pi)^2}[(g_+^L+g_-^L)(g_+^M+g_-^M)\\
		&-\frac{4t^2_Jk^2_xk^2_y}{d^2}(g_+^L-g_-^L)(g_+^M-g_-^M)]
\end{aligned}\end{equation}

Therefore, the longitudinal ($\sigma^{l,\text{int+sj}}_{xx}$) and spin Hall ($\sigma^{s,\text{int+sj}}_{yx}$) conductivities, which include both the intrinsic and side-jump scattering contributions, can be evaluated by replacing one vertex with that dressed by the ladder-type correction in the Kubo formula Eq.~(\ref{eq:Kubo-initial}). For $\sigma^{l,\text{int+sj}}_{xx}$, replace one charge vertex $v_x$ with $\tilde{v}_x$, and for $\sigma^{s,\text{int+sj}}_{yx}$, replace the charge vertex $v_x$ with $\tilde{v}_x$ or the spin vertex $v_{s,y}$ with $\tilde{v}_{s,y}$.

Similarly, we can derive the analytical solutions for $\sigma^{l,\text{int+sj}}_{xx}$ and $\sigma^{s,\text{int+sj}}_{yx}$ in the limits $k\ll\frac{\lambda}{t_J}$ and $\tau\to0$, while neglecting the $RR$ terms ($g^{R}g^{R}$), the interband coherences ($g_{+}g_{-}$, $g_{-}g_{+}$), and the Fermi-sea contribution ($\sigma^{s,II}_{yx}$). These solutions are approximately captured by
\begin{equation}\begin{aligned}\label{eq:analytic-side-jump}
		\sigma^{l,\text{int+sj}}_{xx}\approx&\frac{\pi e^2\tau}{h}\sum_{\gamma=\pm}\int\frac{d^2\bm{k}}{(2\pi)^2}v_x^{\gamma\gamma}\tilde{v}_x^{\gamma\gamma}\delta(E-E_{\bm{k}\gamma})\\
		\sigma^{s,\text{int+sj}}_{yx}\approx&\frac{\pi e^2\tau}{h}\sum_{\gamma=\pm}\int\frac{d^2\bm{k}}{(2\pi)^2}v_{s,y}^{\gamma\gamma}\tilde{v}_x^{\gamma\gamma}\delta(E-E_{\bm{k}\gamma}).
\end{aligned}\end{equation}
where
\begin{equation}\label{eq:analytic-dress-vx}
	\tilde{v}_x^{\gamma\gamma}=2tk_x+\gamma\frac{\lambda^{RA}k_x}{k}
	\ \ \text{with}\ \ 
	\lambda^{RA}=\left\{\begin{array}{ll}
		-\frac{4tE}{\lambda};&E<0\\
		0;&E\ge0
	\end{array}\right.
\end{equation}
is the diagonal component of the dressed charge velocity obtained from Eq.~(\ref{eq:vertex-correction-charge}), corresponding to the renormalized quasiparticle velocity. Plugging Eqs.~(\ref{eq:analytic-vx-sy}) and (\ref{eq:analytic-dress-vx}) into Eq.~(\ref{eq:analytic-side-jump}), we get analytical solutions
\begin{equation}\label{eq:analytic-sxx-sj}
	\sigma^{l,\text{int+sj}}_{xx}
	=\frac{e^2}{h}\frac{2\lambda^2}{n_iV_0^2}\left\{\begin{array}{ll}
		\frac{1}{2}[(\frac{n}{n_0})^4+(\frac{n}{n_0})^2];&n<n_0\\
		(\frac{n}{n_0});&n\ge n_0
	\end{array}\right.
\end{equation}
and
\begin{equation}\label{eq:analytic-syx-sj}
	\sigma^{s,\text{int+sj}}_{yx}
	=\frac{e^2}{h}\frac{2\lambda^2}{n_iV_0^2}\frac{t_J}{t}\left\{\begin{array}{ll}
		\frac{n}{n_0};&n<n_0\\
		1;&n\ge n_0
	\end{array}\right..
\end{equation} 
Contrary to the intrinsic contribution, analysis of the parities of $k_x$ and $k_y$ in Eq.~(\ref{eq:analytic-side-jump}) reveals that the side-jump scattering contribution vanishes in the absence of SOC is absent ($\lambda=0$).

In Fig.~\ref{fig:conductivity}, we show the numerical and analytical results for the spin Hall angle, including contributions from both intrinsic and side-jump scattering mechanisms, $\sigma^{s,\text{int+sj}}_{yx}/\sigma^{l,\text{int+sj}}_{xx}$, as a function of $n/n_0$, compared with results considering only the intrinsic mechanism. We find that below the Dirac point ($n/n_0<1$), the spin Hall angle can be markedly enhanced by the side-jump scattering, while above the Dirac point ($n/n_0>1$), the spin Hall angle shows almost no change with or without side-jump scattering contribution. Additionally, the approximate analytical solutions align well with the numerical results, except in the region extremely close to the band edge, which is dominated by the disorder-induced states.

The renormalized quasiparticle velocity derived analytically, given in Eq.~(\ref{eq:analytic-dress-vx}), is the same as that obtained in the 2D Rashba electron gas without altermagnetic exchange interaction \cite{Brosco2016prl,Chen2020prb}.  This correspondence indicates that the side-jump mechanism effectively suppress backward scattering processes in a similar manner as discussed in Ref~\cite{Brosco2016prl}. In the system considered in this work, such suppression stems from the helicity-driven orthogonality between in-plane spins of states with antiparallel velocities and momenta. As a result, regardless of the Fermi energy's location relative to the Dirac point, backward scattering predominantly occurs between quasiparticle states across different Fermi circles, reversing the quasiparticle velocity's sign without altering momentum direction, as illustrated schematically in Figs.~\ref{fig:band}(e) and (f).

Spin Hall conductivity, unlike longitudinal conductivity which is primarily influenced by the distribution of transport states, is also affected by the polarization of out-of-plane spin $\langle s_z\rangle$. Above the Dirac point, the polarization direction of $\langle s_z\rangle$ in backward scattering states is parallel to that in forward scattering states, whereas it is antiparallel below the Dirac point. As a result, suppressing backward scattering below the Dirac point not only increases the distribution of transport states but also enhances the $\langle s_z\rangle$ polarization. This dual enhancement effect significantly boosts spin Hall conductivity, leading to a pronounced increase in the spin Hall angle.

\begin{figure}
	\flushleft
	\includegraphics[width=0.95\linewidth]{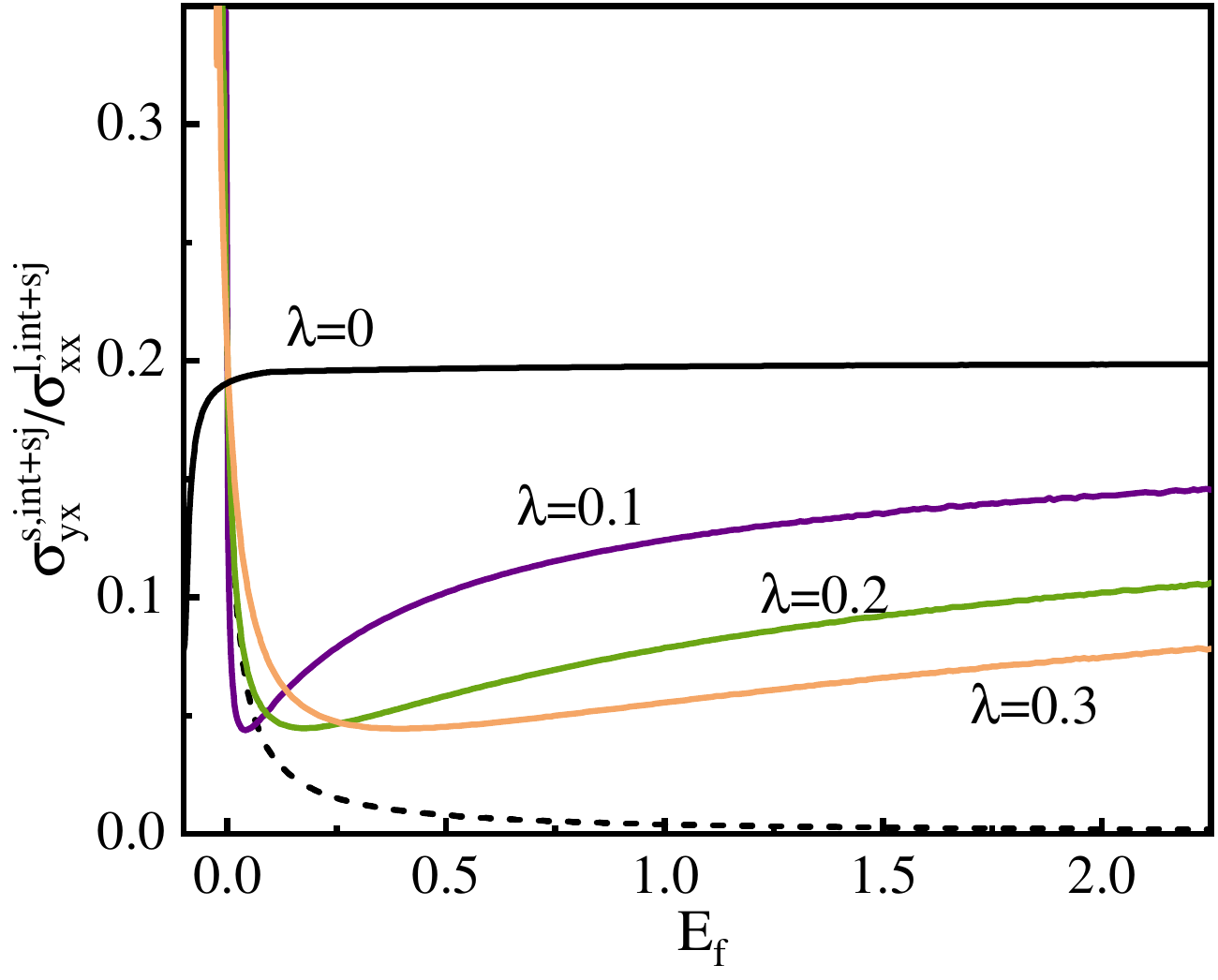}
	\caption{(Color online) $\sigma^{s,\text{int+sj}}_{yx}/\sigma^{l,\text{int+sj}}_{xx}$ as a function of $E_f$ across a broad energy scale. Parameters: $t_J=0.2$, $\lambda=0$ (black), 0.1 (purple), 0.2 (green), 0.3 (orange), and $V_0=0.05$. The dashed line denotes the analytical result given by Eqs.~(\ref{eq:analytic-sxx-sj}-\ref{eq:analytic-syx-sj}).}
	\label{fig:cond-angle-large}
\end{figure}

In Fig.~\ref{fig:cond-angle-large}, we present $\sigma^{s,\text{int+sj}}_{yx}/\sigma^{l,\text{int+sj}}_{xx}$ as a function of $E_f$ across a broad range. The numerical result for $\lambda=0$ is in excellent agreement with the analytical solution in Eq.~(\ref{eq:sigma-lambda0}). However, for $\lambda\ne0$, the analytical and numerical results diverge in the high energy region, due to the assumption that $k\ll\frac{\lambda}{t_J}$ is no longer valid. Upon comparing the result for $\lambda=0$ with those for $\lambda \neq 0$, we find that SOC effect suppresses the spin Hall angle in the region above Dirac point, but this suppression weakens with increasing $E_f$. This trend is linked to a decrease in the in-plane spin component, $\langle s_{\parallel}\rangle$, as shown in Fig.~\ref{fig:band}, which consequently diminishes the influence of SOC-induced helical spin textures on the spin Hall effect.

\textit{Skew scattering.---}
The contribution from skew scattering is evaluated by the Feynman diagrams with a single third order impurity vertex and both external velocity vertices renormalized by ladder-type vertex corrections, as shown in Fig.~\ref{fig:feynman-diagram}(c). Based on these diagrams, the longitudinal ($\sigma^{l,\text{skew}}_{xx}$) and spin Hall ($\sigma^{s,\text{skew}}_{yx}$) conductivities resulting from skew scattering are given by
\begin{equation}\label{eq:skew-total}
	\sigma^{l/s,\text{skew}}_{ij}={\rm Re}(\sigma_{ij}^{l/s,Ia,,\text{skew}}+\sigma_{ij}^{l/s,Ib,,\text{skew}})
\end{equation}
with
\begin{equation}\begin{aligned}\label{eq:skew-sub}
		\sigma_{ij}^{l/s,Ia,\text{skew}}=&\frac{e^2}{h}V_1^3\int\frac{d^2\bm{k}}{(2\pi)^2}{\rm Tr}[\tilde{v}^{AR}_iG_{\bm{k}}^R\tilde{v}^{RA}_j+\tilde{v}^{AR}_i\tilde{v}^{RA}_jG_{\bm{k}}^A]\\
		\sigma_{ij}^{l/s,Ib,\text{skew}}=&-\frac{e^2}{h}V_1^3{\rm Re}\int\frac{d^2\bm{k}}{(2\pi)^2}{\rm Tr}[\tilde{v}^{RR}_iG_{\bm{k}}^R\tilde{v}^{RR}_j+\tilde{v}_i\tilde{v}^{RR}_jG_{\bm{k}}^R]\\
\end{aligned}\end{equation}
where the vertices are set as $\tilde{v}^{LM}_i=\tilde{v}^{LM}_j=\tilde{v}^{LM}_x$ for $\sigma^{l,\text{skew}}_{xx}$, and set as $\tilde{v}^{LM}_i=\tilde{v}^{LM}_{s,y}$ and $\tilde{v}^{LM}_j=\tilde{v}^{LM}_x$ for $\sigma^{s,\text{skew}}_{yx}$.

Numerical calculations reveal that the skew scattering's contribution to these conductivities is negligible compared to those from intrinsic and side-jump scattering mechanisms even for a large $V_1$. This finding contrasts with conventional ferromagnetic systems with Rashba SOC, where skew scattering significantly affects both the anomalous and spin Hall conductivity when third-order disorder correlators are considered \cite{Sinova2004prl,Onoda2006prl,Nunner2008prl}. The key difference arises from the altermagnetic exchange interaction that breaks time-reversal symmetry without inducing asymmetry in the lifetimes of quasiparticle states across the two subbands.

\textit{Discussion and conclusion.---} In conclusion, we explore the charge-to-spin conversion in 2D Rashba altermagnets by calculating the spin Hall angle using quantum linear response theory. We find that the side-jump mechanism can markedly enhance the spin Hall conductivity below the band crossing point, thereby facilitating the achievement of a giant spin Hall angle. Through analyzing the in-plane and out-of-plane spin textures, we demonstrate that side-jump mechanism produces a dual enhancement effect on the generation of spin current in this region: suppressing backward scattering and increasing spin polarization. Furthermore, we find that away from the Dirac point, the influence of SOC-induced helical spin textures on the spin Hall effect is diminished due to a reduction in the in-plane spin component.

Before ending, several remarks are in order:
First, our findings propose a novel strategy for manipulating charge-to-spin conversion through the sophisticated control of in-plane and out-of-plane spin textures, providing insights for further high-efficiency conversion in altermagnets. 
Second, there has been a significant increase in reports on 2D altermagnetic materials characterized by marked anisotropic spin-splitting and strong spin-orbit coupling \cite{Liu2024afm,Sodequist2024apl,Parfenov2024arxiv,Mazin2023arxiv,Wu2024nc}, which can be suitable platforms to achieve our idea. 
Third, in contrast to typical ferromagnetic Rashba systems, the skew scattering mechanism and topology-related intrinsic mechanism are almost negligible in this system, both of which are attributable to unique time-reversal symmetry breaking in altermagnets. Therefore, our research is expected to catalyze further investigations into the uniqueness of time-reversal symmetry breaking in altermagnetic materials, potentially expanding the scope of applications in spintronics.

\begin{acknowledgments}
This work is supported by National Science Foundation of China (92165102), and the foundation from Westlake University. This work was supported by ``Pioneer" and ''Leading Goose" R\&D Program of Zhejiang (2022SDXHDX0005), the Key R\&D Program of Zhejiang Province (2021C01002).
\par
\end{acknowledgments}

\begin{appendix}

\end{appendix}


\begin{thebibliography}{99}
	
\bibitem{Smejkal2022prx2} L, Smejkal, J. Sinova, and T. Jungwirth,
Emerging Research Landscape of Altermagnetism,
\href{https://journals.aps.org/prx/abstract/10.1103/PhysRevX.12.040501}
{Phys. Rev. X \textbf{12}, 040501 (2022).} 

\bibitem{Smejkal2022nrm} L. Smejkal, A. H. MacDonald, J. Sinova, S. Nakatsuji, and T. Jungwirth,
Anomalous Hall antiferromagnets, 
\href{https://www.nature.com/articles/s41578-022-00430-3}
{Nat. Rev. Mater. \textbf{7}, 482 (2022).}

\bibitem{Ouassou2023prl} J. A. Ouassou, A. Brataas, and J. Linder,
dc Josephson Effect in Altermagnets,
\href{https://journals.aps.org/prl/abstract/10.1103/PhysRevLett.131.076003}
{Phys. Rev. Lett. \textbf{131}, 076003 (2023).}

\bibitem{Smejkal2022prx1} L. Smejkal, J. Sinova, and T. Jungwirth,
Beyond Conventional Ferromagnetism and Antiferromagnetism: A Phase with Nonrelativistic Spin and Crystal Rotation Symmetry,
\href{https://journals.aps.org/prx/abstract/10.1103/PhysRevX.12.031042}
{Phys. Rev. X \textbf{12}, 031042 (2022).}



\bibitem{Bai2023prl} H. Bai, Y. C. Zhang, Y. J. Zhou, P. Chen, C. H. Wan, L. Han, W. X. Zhu, S. X. Liang, Y. C. Su, X. F. Han, F. Pan, and C. Song,
Efficient Spin-to-Charge Conversion via Altermagnetic Spin Splitting Effect in Antiferromagnet RuO$_2$,
\href{https://journals.aps.org/prl/abstract/10.1103/PhysRevLett.130.216701}
{Phys. Rev. Lett. \textbf{130}, 216701 (2023)}.

\bibitem{Krempasky2024nature} J. Krempask\'y, L. \u{S}mejkal, S. W. D'Souza, M. Hajlaoui, G. Springholz, K. Uhli\u{r}ov\'a, F. Alarab, P. C. Constantinou, V. Strocov, D. Usanov, W. R. Pudelko, R. Gonz\'alez-Hern\'andez, A. Birk Hellenes, Z. Jansa, H. Reichlov\'a, Z. \u{S}ob\'a\u{n}, R. D. Gonzalez Betancourt, P. Wadley, J. Sinova, D. Kriegner, J. Min\'ar, J. H. Dil and T. Jungwirth,
Altermagnetic lifting of Kramers spin degeneracy,
\href{https://www.nature.com/articles/s41586-023-06907-7#citeas}
{Nature \textbf{626}, 517–522 (2024).}

\bibitem{Smejkal2020sa} L. \u{S}mejkal, R. G.-Hern\'andez, T. Jungwirth, J. Sinova,
Crystal time-reversal symmetry breaking and spontaneous Hall effect in collinear antiferromagnets,
\href{https://www.science.org/doi/10.1126/sciadv.aaz8809}
{Sci. Adv. \textbf{6}, eaaz8809 (2020).}

\bibitem{Osumi2024prb} T. Osumi, S. Souma, T. Aoyama, K. Yamauchi, A. Honma, K. Nakayama, T. Takahashi, K. Ohgushi, and T. Sato,
Observation of a giant band splitting in altermagnetic MnTe,
\href{https://journals.aps.org/prb/abstract/10.1103/PhysRevB.109.115102}
{Phys. Rev. B \textbf{109}, 115102 (2024).}


\bibitem{Hernandez2021prl} R. G.-Hern\'andez, L. \u{S}mejkal, K. V\'yborn\'y, Y. Yahagi, J. Sinova, T. Jungwirth, and J. \u{Z}elezn\'y,
Efficient Electrical Spin Splitter Based on Nonrelativistic Collinear Antiferromagnetism,
\href{https://journals.aps.org/prl/abstract/10.1103/PhysRevLett.126.127701}
{Phys. Rev. Lett. \textbf{126}, 127701 (2021).}

\bibitem{Bai2022prl} H. Bai, L. Han, X. Y. Feng, Y. J. Zhou, R. X. Su, Q. Wang, L. Y. Liao, W. X. Zhu, X. Z. Chen, F. Pan, X. L. Fan, and C. Song,
Observation of Spin Splitting Torque in a Collinear Antiferromagnet RuO$_2$,
\href{https://journals.aps.org/prl/abstract/10.1103/PhysRevLett.128.197202}
{Phys. Rev. Lett. \textbf{128}, 197202 (2022).}

\bibitem{Naka2021prb} Makoto Naka, Yukitoshi Motome, and Hitoshi Seo,
Perovskite as a spin current generator,
\href{https://journals.aps.org/prb/abstract/10.1103/PhysRevB.103.125114}
{Phys. Rev. B \textbf{103}, 125114 (2021).}


\bibitem{Naka2019nc} M. Naka, S. Hayami, H. Kusunose, Y. Yanagi, Y. Motome, and H. Seo,
Spin current generation in organic antiferromagnets,
\href{https://www.nature.com/articles/s41467-019-12229-y}
{Nat. Commun. \textbf{10}, 4305 (2019).}

\bibitem{Li2024arxiv} Z. Li, Z. Zhang, X. Lu, Y. Xu,
Spin Splitting in Altermagnetic RuO$_2$ Enables Field-free Spin-Orbit Torque Switching via Dominant Out-of-Plane Spin Polarization,
\href{https://arxiv.org/abs/2407.07447}
{arXiv:2407.07447}

\bibitem{Ma2021nc} Hai-Yang Ma, Mengli Hu, Nana Li, Jianpeng Liu, Wang Yao, Jin-Feng Jia, and Junwei Liu,
Multifunctional antiferromagnetic materials with giant piezomagnetism and noncollinear spin current,
\href{https://www.nature.com/articles/s41467-021-23127-7}
{Nat. Commun. \textbf{12}, 2846 (2021).}

\bibitem{Karube2022prl} S. Karube, T. Tanaka, D. Sugawara, N. Kadoguchi, M. Kohda, and J. Nitta,
Observation of Spin-Splitter Torque in Collinear Antiferromagnetic RuO$_2$,
\href{https://journals.aps.org/prl/abstract/10.1103/PhysRevLett.129.137201}
{Phys. Rev. Lett. \textbf{129}, 137201 (2022).}

\bibitem{Reimers2024nc} S. Reimers, L. Odenbreit,L. \u{S}mejkal, V. N. Strocov, P. Constantinou, A. B. Hellenes, R. J. Ubiergo, W. H. Campos, V. K. Bharadwaj,A. Chakraborty, T. Denneulin, W. Shi, R. E. Dunin-Borkowski, S. Das, M. Kl\"aui, J. Sinova1, and M. Jourdan, Direct observation of altermagnetic band splitting in CrSb thin films,
\href{https://doi.org/10.1038/s41467-024-46476-5}
{Nat. Commun. \textbf{15}, 2116 (2024).}

\bibitem{Reichlova2024nc} H. Reichlova, R. L. Seeger, R. G-Hern\'andez, I. Kounta, R. Schlitz, D. Kriegner, P. Ritzinger, M. Lammel, M. Leivisk\"a, A. B. Hellenes, K. Olejnik, Vaclav Pet\u{u}i\u{c}ek, P. Dole\u{z}al, L. Horak, E. Schmoranzerova, A. Badura, S. Bertaina, A. Thomas, V. Baltz, L. Michez, J. Sinova, S. T. B. Goennenwein, T. Jungwirth, and L. \u{S}mejkal, Observation of a spontaneous anomalous Hall response in the Mn$_5$Si$_3$ d-wave altermagnet candidate,
\href{https://doi.org/10.1038/s41467-024-48493-w}
{Nat. Commun. \textbf{15}, 4961 (2024).}

\bibitem{Sasabe2023prl} N. Sasabe, M. Mizumaki, T. Uozumi, and Y. Yamasaki,
Ferroic Order for Anisotropic Magnetic Dipole Term in Collinear Antiferromagnets of $(t_{2g})^4$ System,
\href{https://journals.aps.org/prl/abstract/10.1103/PhysRevLett.131.216501}
{Phys. Rev. Lett. \textbf{131}, 216501 (2023).}

\bibitem{Seo2021cp} S. Seo, S. Hayami, Y. Su, S. M. Thomas, F. Ronning, E. D. Bauer, J. D. Thompson, S-Z. Lin, and P. F. S. Rosa,
Spin-texture-driven electrical transport in multi-Q antiferromagnets,
\href{https://www.nature.com/articles/s42005-021-00558-8}
{Commun. Phys. \textbf{4}, 58 (2021).}

\bibitem{Hua2024prb} J. Hua, Z. F. Wang, W. Zhu, and W. Chen,
Chirality-2 fermion induced anti-Klein tunneling in a two-dimensional checkerboard lattice,
\href{https://journals.aps.org/prb/abstract/10.1103/PhysRevB.109.115429}
{Phys. Rev. B \textbf{109}, 115429 (2024).}

\bibitem{Gorini2008prb} C. Gorini, P. Schwab, M. Dzierzawa, and R. Raimondi,
Spin polarizations and spin Hall currents in a two-dimensional electron gas with magnetic impurities,
\href{https://journals.aps.org/prb/abstract/10.1103/PhysRevB.78.125327}
{Phys. Rev. B \textbf{78}, 125327 (2008).}

\bibitem{Chen2024prb} Y. Chen, Z. Z. Du, H-Z. Lu, and X. C. Xie,
Intrinsic spin-orbit torque mechanism for deterministic all-electric switching of noncollinear antiferromagnets,
\href{https://journals.aps.org/prb/abstract/10.1103/PhysRevB.109.L121115}
{Phys. Rev. B \textbf{109}, L121115 (2024)}.

\bibitem{Smejkal2022prx3} L. \u{S}mejkal, A. B. Hellenes, R. Gonz\'alez-Hern\'andez, J. Sinova, and T. Jungwirth,
Giant and Tunneling Magnetoresistance in Unconventional Collinear Antiferromagnets with Nonrelativistic Spin-Momentum Coupling,
\href{https://journals.aps.org/prx/abstract/10.1103/PhysRevX.12.011028}
{Phys. Rev. X \textbf{12}, 011028 (2022).}

\bibitem{Cui2023prb} Q. Cui, Y. Zhu, X. Yao, P. Cui, and H. Yang,
Giant spin-Hall and tunneling magnetoresistance effects based on a two-dimensional nonrelativistic antiferromagnetic metal,
\href{https://journals.aps.org/prb/abstract/10.1103/PhysRevB.108.024410}
{Phys. Rev. B \textbf{108}, 024410 (2023).}

\bibitem{Brosco2016prl} V. Brosco, L. Benfatto, E. Cappelluti, and C. Grimaldi,
Unconventional dc Transport in Rashba Electron Gases,
\href{https://journals.aps.org/prl/abstract/10.1103/PhysRevLett.116.166602}
{Phys. Rev. Lett. \textbf{116}, 166602 (2016).}

\bibitem{Sinova2015rmp} J. Sinova, S. O. Valenzuela, J. Wunderlich, C. H. Back, and T. Jungwirth,
Spin Hall effects,
\href{https://journals.aps.org/rmp/abstract/10.1103/RevModPhys.87.1213}
{Rev. Mod. Phys. \textbf{87}, 1213 (2015).}

\bibitem{Tse2006prb} W-K. Tse and S. Das Sarma,
Intrinsic spin Hall effect in the presence of extrinsic spin-orbit scattering
\href{https://journals.aps.org/prb/abstract/10.1103/PhysRevB.74.245309}
{Phys. Rev. B \textbf{74}, 245309 (2006).}

\bibitem{Nunner2008prl} T. S. Nunner, G. Zar\'and, and F. von Oppen,
Anomalous Hall Effect in a Two Dimensional Electron Gas with Magnetic Impurities,
\href{https://journals.aps.org/prl/abstract/10.1103/PhysRevLett.100.236602}
{Phys. Rev. Lett. \textbf{100}, 236602 (2008)}.

\bibitem{Borunda2007prl} M. Borunda, T. S. Nunner, T. Luck, N. A. Sinitsyn, C. Timm, J. Wunderlich, T. Jungwirth, A. H. MacDonald, and J. Sinova,
Absence of Skew Scattering in Two-Dimensional Systems: Testing the Origins of the Anomalous Hall Effect,
\href{https://journals.aps.org/prl/abstract/10.1103/PhysRevLett.99.066604}
{Phys. Rev. Lett. \textbf{99}, 066604 (2007)}.

\bibitem{Nunner2007prb} T. S. Nunner, N. A. Sinitsyn, M. F. Borunda, V. K. Dugaev, A. A. Kovalev, Ar. Abanov, C. Timm, T. Jungwirth, Jun-ichiro Inoue, A. H. MacDonald, and J. Sinova,
Anomalous Hall effect in a two-dimensional electron gas,
\href{https://journals.aps.org/prb/abstract/10.1103/PhysRevB.76.235312}
{Phys. Rev. B \textbf{76}, 235312 (2007)}.

\bibitem{Kovalev2009prb} A. A. Kovalev, Y. Tserkovnyak, K. V\'yborn\'y, and J. Sinova,
Transport theory for disordered multiple-band systems: Anomalous Hall effect and anisotropic magnetoresistance,
\href{https://journals.aps.org/prb/abstract/10.1103/PhysRevB.79.195129}
{Phys. Rev. B \textbf{79}, 195129 (2009).}

\bibitem{Yang2011prb} S. A. Yang, H. Pan, Y. Yao, and Q. Niu,
Scattering universality classes of side jump in the anomalous Hall effect,
\href{https://journals.aps.org/prb/abstract/10.1103/PhysRevB.83.125122}
{Phys. Rev. B \textbf{83}, 125122 (2011).}

\bibitem{Chen2020prb} W. Chen, C. Xiao, Q. Shi, and Q. Li,
Spin-orbit related power-law dependence of the diffusive conductivity on the carrier density in disordered Rashba two-dimensional electron systems,
\href{https://journals.aps.org/prb/abstract/10.1103/PhysRevB.101.020203}
{Phys. Rev. B \textbf{101}, 020203(R) (2020).}








\bibitem{Sinova2004prl} J. Sinova, D. Culcer, Q. Niu, N. A. Sinitsyn, T. Jungwirth, and A. H. MacDonald,
Universal Intrinsic Spin Hall Effect,
\href{https://journals.aps.org/prl/abstract/10.1103/PhysRevLett.92.126603}
{Phys. Rev. Lett. \textbf{92}, 126603 (2004).}

\bibitem{Onoda2006prl} S. Onoda, N. Sugimoto, and N. Nagaosa,
Intrinsic Versus Extrinsic Anomalous Hall Effect in Ferromagnets,
\href{https://journals.aps.org/prl/abstract/10.1103/PhysRevLett.97.126602}
{Phys. Rev. Lett. \textbf{97}, 126602 (2006).}



\bibitem{Liu2024afm} Q. Liu, J. Kang, P. Wang, W. Gao, Y. Qi, J. Zhao, and X. Jiang,
Inverse Magnetocaloric Effect in Altermagnetic 2D Non-van der Waals FeX (X = S and Se) Semiconductors,
\href{https://doi.org/10.1002/adfm.202402080}
{Adv. Funct. Mater. 2402080 (2024).}

\bibitem{Sodequist2024apl} J. S\o dequist, T. Olsen,
Two-dimensional altermagnets from high throughput computational screening: Symmetry requirements, chiral magnons, and spin-orbit effects,
\href{https://doi.org/10.1063/5.0198285}
{Appl. Phys. Lett. \textbf{124}, 182409 (2024).}

\bibitem{Parfenov2024arxiv} O. E. Parfenov, D. V. Averyanov, I. S. Sokolov, A. N. Mihalyuk, O. A. Kondratev, A. N. Taldenkov, A. M. Tokmachev, and V. G. Storchak,
Pushing an Altermagnet to the Ultimate 2D Limit: Evidence of Symmetry Breaking in Monolayers of GdAlSi
\href{https://doi.org/10.48550/arXiv.2406.07172}
{arXiv:2406.07172}

\bibitem{Mazin2023arxiv} I. Mazin, R. G-Hern\'andez, L. \u{S}mejkal,
Induced Monolayer Altermagnetism in MnP(S,Se)$_3$ and FeSe,
\href{https://doi.org/10.48550/arXiv.2309.02355}
{arXiv:2309.02355}

\bibitem{Wu2024nc} Y. Wu, L. Deng, X. Yin, J. Tong, F. Tian, and X. Zhang,
Valley-Related Multipiezo Effect and Noncollinear Spin Current in an Altermagnet Fe$_2$Se$_2$O Monolayer,
\href{https://doi.org/10.1021/acs.nanolett.4c02554}
{Nano. Lett. \textbf{24}, 10534-10539 (2024).}











\end{thebibliography}
\end{document}